\documentclass[onecolumn, 11pt]{revtex4} 
\usepackage{amsfonts}
\usepackage{amsmath}
\usepackage{amssymb}
\usepackage{graphicx}

\begin{document}
\title[Decoherence and the (non)emergence of classicality]{Decoherence and the (non)emergence of classicality}

\author{Steven Weinstein}

\email{sweinstein@perimeterinstitute.ca; sw@uwaterloo.ca}

\affiliation{Perimeter Institute for Theoretical Physics, 31 Caroline St N, Waterloo ON N2L\ 2Y5 Canada}


\begin{abstract}
We consider the claim that decoherence explains the emergence of classicality
in quantum systems, and conclude that it does not. We show that, given a
randomly chosen universe composed of a variety of subsystems, some of which
are macroscopic and subject to decoherence-inducing interactions, and some of
which are microscopic, the macroscopic subsystems will not display any
distinctively classical behavior. Therefore, a universe in which macroscopic
and microscopic do display distinct behavior must be in a very special, highly
nongeneric quantum state.

\end{abstract}
\received{October 24, 2008}

\maketitle

\section{Introduction}

Over the last four decades \cite{Zeh70}, the study of decoherence has begun to
shed light on the effects of the interaction of open quantum systems with
their environments.  It has been shown that, for some interesting model
systems, certain pure states, sometimes called \emph{pointer states}%
\ \cite{Zur81}\ \cite{Zur82} survive interaction with the environment without
losing their coherence or purity, while superpositions of these states lose
coherence in such a way that the result is an incoherent, improper
\emph{mixture }of such states which is subsequently approximately stable.

The fact that macroscopic subsystems interacting with an appropriate
environment can be seen to exhibit decoherence in a preferred basis,  along
with the fact that the basis in question often corresponds to a
paradigmatically classical observable such as position, has led to claims that
\textquotedblleft the classical structure of phase space emerges from the
quantum Hilbert space in the appropriate limit\textquotedblright%
\ \cite{Zur03}; that \textquotedblleft the appearance of classicality is
therefore grounded in the structure of the physical laws governing the
system-system environment interactions\textquotedblright\cite{Schl07}; and
that \textquotedblleft there are strong signs that the transition [from
quantum to classical] can be understood as something that emerges quite
naturally and inevitably from quantum theory\textquotedblright\cite{Ball08}.
Other, similar claims lie ready to hand \cite{Zur91}\cite{Joos03}. In this
paper, we show that the properties of generic microscopic subsystems of a
quantum-mechanical universe are kinematically and dynamically
indistinguishable from the properties of generic macroscopic subsystems, and
thereby show that decoherence does not explain the emergence of (quasi)classicality.

\section{Decoherence, einselection, and quasiclassicality}

Consider a subsystem $S$ with (pure) state $\psi^{S}$ interacting with an
environment $E$ with state $\psi^{E}$. If the subsystem is sufficiently
macroscopic, and if the Hamiltonian governing the combined evolution of
subsystem and environment is appropriate, then the environment as a whole acts
as a kind of measuring device, in that the effective state of the environment
(given by its reduced density matrix) will reliably become correlated with
certain subsystem observables. Which properties of the system are
\textquotedblleft measured\textquotedblright\ by the environment -- which
observables (if any) become nontrivially correlated -- will depend on the
Hamiltonian, including the self-Hamiltonians of system and of environment
\cite{DHMM05}. Eigenstates of the subsystem observables in question, the
pointer states, will be stable or approximately stable under such
measurement-like interactions, while arbitrary \emph{superpositions} of
pointer states will evolve into improper \emph{mixtures} of those states as a
result of the environment's correlation with the pointer observable. The
tendency for the reduced density matrix of the  subsystem to be driven into a
small subset of the available states by the environment is called
\emph{einselection}, short for \emph{environment induced superselection}
\cite{Zur82}\cite{Zur03}.

Decoherence, then, refers to the process by which pure states lose their
coherence, and more particularly to the way in which sufficiently macroscopic
subsystems lose their coherence in a way characterized by einselection.
Subsystems which undergo einselection said to be quasiclassical (sometimes
simply \textquotedblleft classical\textquotedblright) in virtue of their
stability and predictability; they not only lose coherence -- so do many
nonclassical, microscopic systems -- but they do so in a predictable way, and
evolve stably thereafter. (The qualification \textquotedblleft
quasi\textquotedblright\ is in place because an improper mixture has no direct
classical analog, and because einselection is never exact and is subject to
Poincare recurrences.)

\subsection{Example:\ Central spin model}

Consider for example the so-called central spin model \cite{Zur82}, in which
one contemplates a system consisting of $N+1$ two-level systems, $N$ of which
are coupled to a central spin $S$ via the Hamiltonian%
\[
\hat{H}=\frac{1}{2}\hat{\sigma}_{z}\otimes\left(  \sum\limits_{i=1}^{N}%
g_{i}\hat{\sigma}_{z}^{(i)}%
\bigotimes_{i' \not= i}%
\hat{I}_{i^{\prime}}\right)
\]
where $\hat{I}_{i}$ is the identity operator for the $i$'th system. (Here
there is no macroscopic/microscopic distinction; rather, the distinctive
dynamical role of the central spin singles it out as special.) An initial pure
state of the form
\[
\psi=\alpha\left\vert +z\right\rangle \left\vert E_{0}\right\rangle
+\beta\left\vert -z\right\rangle \left\vert E_{0}\right\rangle
\]
will, via the unitary evolution $U(t)=e^{-i\hat{H}t}$ generated by this
Hamiltonian, evolve toward an entangled state $\psi(t)=\alpha\left\vert
+z\right\rangle \left\vert E_{+}(t)\right\rangle +\beta\left\vert
-z\right\rangle \left\vert E_{-}(t)\right\rangle $. After a sufficient amount
of time $t_{d}$ has passed, $\left\langle E_{+}|E_{-}\right\rangle \approx0$,
and the reduced density matrix of the central spin will be well-approximated
by
\[
\rho^{S}=\alpha^{2}\left\vert +z\right\rangle \left\langle +z\right\vert
+\beta^{2}\left\vert -z\right\rangle \left\langle -z\right\vert \text{.}%
\]
One can represent this evolution on the Bloch sphere as the evolution of
initially pure states of the central qubit (the surface of the sphere),
evolving, modulo extremely unlikely Poincare-type fluctuations, toward a
narrow ellipse along the $z$ axis:%

\begin{center}
\includegraphics[
height=2.0289in,
width=2.0289in
]%
{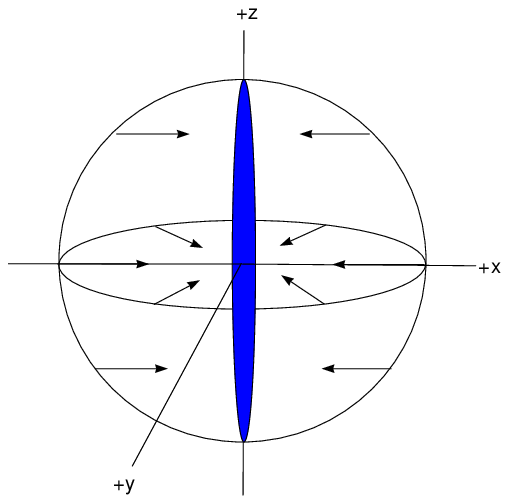}%
\\
Evolution of the central spin
\end{center}

\noindent Though it is no surprise that a subsystem should lose coherence upon
interaction with an environment, and thus move away from the surface of the
Bloch sphere, we have here in addition the phenomenon of einselection, in
which the loss of coherence is in a preferred direction. In particular, the
loss of purity is proportional to the angle with the $z$ axis, with the
pointer states $\left\vert +z\right\rangle $ and $\left\vert -z\right\rangle $
suffering no loss whatsoever. 

\section{Properties of generic subsystems}

Consider a large system described by an arbitrary pure state $\rho=\left\vert
\psi\right\rangle \left\langle \psi\right\vert $ defined over a Hilbert space
$\mathcal{H}$ which can be decomposed into a tensor-product of Hilbert spaces
$\mathcal{H}_{i}$, one of which corresponds to the subsystem $S$ of interest
and the rest of which collectively correspond to the environment $E$. Thus
\begin{equation}
\mathcal{H=H}_{S}\mathcal{\otimes H}_{2}\mathcal{\otimes H}_{3}%
\mathcal{\otimes}...\otimes\mathcal{H}_{k}\text{.}\label{Hilbert}%
\end{equation}
The state $\left\vert \psi\right\rangle $ $\in\mathcal{H}$ with density matrix
$\rho=\left\vert \psi\right\rangle \left\langle \psi\right\vert $ is then just
a vector in this space, and the state of $S$, $\rho_{S}=Tr_{E}(\rho)$, is just
the partial trace of $\rho$ over the environment $E$. The claim we are
assessing is that, given the appropriate Hamiltonian, subsystems $S$ which are
appropriately macroscopic (or possessed of other properties which make them
suitable candidates for quasiclassicality) undergo einselection and
subsequently evolve in such a way as to exhibit stability in a preferred basis
of pointer states, thus dynamically distinguishing them from the states of
microscopic systems. We now show that this is not the case.

\subsection{Kinematics}

For simplicity, we restrict attention to finite-dimensional Hilbert spaces, as
in eqn. (\ref{Hilbert}). The physical states are the unit vectors in this
space, and an unbiased probability measure on these states is naturally
defined via the Haar measure over $SU(n)$, where $n=\dim(\mathcal{H})$. This
measure has the desirable property that no particular basis is privileged. The
ensemble of equiprobable states is given by the density matrix \
\begin{equation}
\Omega=%
{\displaystyle\sum\limits_{i=1}^{n}}
\frac{1}{n}\left\vert \psi_{i}\right\rangle \left\langle \psi_{i}\right\vert
\end{equation}
where the $\left\vert \psi_{i}\right\rangle $ constitute an orthonormal basis
for the Hilbert space. The rotational invariance of the measure means that
this density matrix looks the same in any basis; it is simply a multiple of
the identity, with $Tr(\Omega)=1$.

Let us consider what one can say about the typical properties of our arbitrary
pure state $\rho=\left\vert \psi\right\rangle \left\langle \psi\right\vert $.
\ In particular, we would like to ask about the properties of the reduced
density matrices $\rho_{S}=Tr_{E}(\rho)$ of subsystems $S$ of dimension
$m<<n$. Popescu \emph{et al} \cite{Pop06} show that, for the vast majority of
states, $\rho_{S}\approx Tr_{E}(\Omega)\equiv\Omega_{S}$. \ More specifically,
they use Levy's lemma to exhibit a bound
\begin{equation}
0<\left\langle D(\rho_{S},\Omega_{S})\right\rangle \leq\frac{m}{2}\sqrt
{\frac{1}{n}}\label{Disting}%
\end{equation}
on the average distinguishability of $\rho_{S}$ and $\Omega_{S}$, where
$D(\rho,\Omega):=\frac{1}{2}Tr\sqrt{(\rho-\Omega)^{\dag}(\rho-\Omega)}%
$.\ Since\ $D(\rho,\Omega)\ $is by definition always positive, the smallness
of the \emph{average} distinguishability $\left\langle D(\rho_{S},\Omega
_{S})\right\rangle $ implies that, for the vast majority of subsystem density
matrices $\rho_{S}$, $D(\rho_{S},\Omega_{S})$ is small as well. As long as the
dimension $m$ of a subsystem is much smaller than the total Hilbert space
dimension $n$, then for almost all states $\rho$, the properties of $S$ will
be essentially indistinguishable from the ensemble average for $S$, which is
$\Omega_{S}$.\ Since $\Omega$ is a multiple of the identity, so too is
$\Omega_{S}=Tr_{E}(\Omega)$, which means that an arbitrarily chosen small
subsystem \emph{behaves as if it were described by a maximally mixed state},
which is in turn to say it behaves randomly with respect to any choice of
observable. \emph{A fortiori, }generic subsystems exhibit no preferred basis.
E.g., the density of states for the central spin has the form $\varrho
(r)\approx(1-r^{2})^{n-2}$, where $n$ is the dimension of the environment and
where $0\leq r\leq1$ is the radius of the Bloch sphere \cite{ZS01}.

\subsection{Dynamics}

Of course the decoherence program does not insist that generic states
$\rho_{S}$ have any special properties.\ Rather, the suggestion is that
systems which are prone to decoherence and einselection \emph{evolve} in such
a way that they \emph{become} quasiclassical. We now show that the vast
majority of subsystems, microscopic or macroscopic, begin in and remain in a
nearly maximally-mixed state. If such states are deemed quasiclassical (in
virtue of their stability), then almost all subsystems are quasiclassical. If
they are not deemed quasiclassical (because they have maximal von Neumann
entropy and carry no information), then almost no states, macroscopic or
microscopic, are quasiclassical.\ In either case, there is no difference
between the behavior of a typical macroscopic and a typical microscopic subsystem.

Suppose that after some time $t_{d}$, the density matrices $\rho_{S}^{\prime
}=Tr_{E}(e^{iHt_{d}}\rho e^{-iHt_{d}})$ for the subsystems $S$ have become
significantly distinct, on average, from the maximally mixed state $\Omega
_{S}$. This implies that $\left\langle D(\rho_{S}^{\prime},\Omega
_{S})\right\rangle $ can exceed the initial bound $\frac{m}{2}\sqrt{\frac
{1}{n}}$. But it is readily seen that this \emph{cannot} be the case, no
matter what form the Hamiltonian takes. For just as the classical Liouville
theorem informs us that a uniform distribution on the phase space will remain
uniform over time, the quantum analogue of the theorem tells us that our
initially uniform distribution $\Omega$ will remain constant under the unitary
evolution $U=e^{iHt_{d}}$ \cite{Tol80}. In particular,%
\begin{equation}
\Omega\overset{U(H,t_{d})}{\rightarrow}\Omega^{\prime}=\Omega\text{ .}%
\end{equation}
Since $\Omega_{S}=Tr_{E}(\Omega)=Tr_{E}(\Omega^{\prime})$, we have%
\begin{equation}
\Omega_{S}\overset{}{\rightarrow}\Omega_{S}^{\prime}=\Omega_{S}\text{.}
\label{Stationary}%
\end{equation}
(The evolution of the density matrices for the subsystems is not unitary, of
course, but it is induced by the unitary evolution for the system as a whole.)
\ We know that, as in equation (\ref{Disting}),
\begin{equation}
0<\left\langle D(\rho_{S}^{\prime},\Omega_{S}^{\prime})\right\rangle \leq
\frac{m}{2}\sqrt{\frac{1}{n}}\text{,}%
\end{equation}
and using\ equation (\ref{Stationary}) we can write%
\begin{equation}
0<\left\langle D(\rho_{S}^{\prime},\Omega_{S})\right\rangle \leq\frac{m}%
{2}\sqrt{\frac{1}{n}}\text{.}%
\end{equation}
Thus the bound on the average distinguishability of $\rho_{S}$ from maximally
mixed $\Omega_{S}$ is \emph{constant} in time, and so there is no
evolution\ toward a preferred basis.

If the emergence of quasiclassicality means the emergence of a preferred basis
for generic, suitably macroscopic subsystems, then decoherence does not
explain the emergence of quasiclassicality. If on the other hand it simply
means stability of generic subsystems, then we certainly have that, for both
microscopic and macroscopic subsystems, since the maximally mixed states which
dominate the ensemble are stable over time. However, this sort of stability
has nothing to do with the choice of Hamiltonian, and it is not specific to
macroscopic subsystems. In fact, from an information-theoretic standpoint, it
would seem to be a rather trivial kind of stability. Given that it means that
the subsystem is overwhelmingly likely to be in a state of maximal (von
Neumann) entropy, the stability of the state of the subsystem simply reflects
the fact that it is a subsystem about which one knows nothing and about which
one continues to know nothing over time.

What decoherence \emph{does} tell us is that \emph{if} we have a subsystem
with suitable properties which is \emph{not }in a maximally mixed state, which
is interacting in an appropriate way with its environment, it will evolve into
a stable, quasiclassical state if it is not already in one. Thus for the
central spin in the example above, we can say with a high degree of certainty
that the rare spin which begins in a state away from the origin of the Bloch
sphere will evolve to, and remain in, a state which is well-approximated by an
improper mixture of $\left\vert +z\right\rangle $ and $\left\vert
-z\right\rangle $ states. But note, too, that the time-reversibility of the
dynamics tells us that with overwhelming probability, this rare spin state
must also have come \emph{from} such a mixture in the moments immediately
prior. Thus decoherence explains how very special, highly non-generic states
manage to maintain their quasiclassicality. It does not explain how
quasiclassicality \textquotedblleft emerges\textquotedblright\ for generic subsystems.

\section{Discussion}

The main feature that emerges from our discussion is that the states of
generic subsystems are stable, be they macroscopic or microscopic, and thus
that the distinctively classical behavior of \emph{macroscopic} subsystems is
still in need of explanation. Certainly, atypical macroscopic subsystems may
exhibit a dynamical behavior which is distinct from the dynamical behavior of
their microscopic counterparts. \ But such subsystems are far from generic.

The fact that distinctive behavior only emerges for subsystems starting in
highly atypical states suggests a connection with the second law of
thermodynamics. \ Indeed, it has been suggested, in the context of the
decoherent \emph{histories }framework \cite{GH90b}\ that the emergence of
classicality is indeed a function of very special initial conditions, and that
the same constraints on initial conditions which yield an arrow of time (i.e.
a monotonic increase in thermodynamic entropy) are those that lead to the
emergence of classical behavior \cite{Har94}\cite{Har08}. This is clearly a
matter worthy of further investigation, and furthermore a matter which is
likely to shed light on the relation between the open-systems decoherence
models studied here and the closed system decoherent-histories models studied elsewhere.

Thanks to Miles Blencowe, Cliff Burgess, Paul Davies, Jim Hartle, Lee Smolin,
Max Schlosshauer, Dieter Zeh, and Wojiech Zurek for comments on an early
draft, and to Robin\ Blume-Kohout, Adrian Kent, Owen Maroney and Jos Uffink
for interesting and enjoyable discussions.

\end{document}